\documentclass[twoside,leqno,twocolumn]{article}  
\usepackage{ltexpprt} 
\usepackage{epsfig}
\usepackage{algorithm}
\usepackage{algpseudocode}
\usepackage{amssymb}
\usepackage{amsmath}
\usepackage{ifpdf}
\usepackage{url}

\algnotext{EndFor}
\algnotext{EndWhile}
\algnotext{EndIf}
\algnotext{EndFunction}

\newcommand{\ma}{{MA$_k$}}

\begin{document}

\title{\Large Crowdsourced Task Routing via Matrix Factorization}
\author{Hyun Joon Jung and Matthew Lease\thanks{School of Information, University of Texas at Austin. Correspondence: \texttt{hyunJoon@utexas.edu}, \texttt{ml@ischool.utexas.edu}.}}

\date{}

\maketitle


\begin{abstract}
We describe methods to predict a crowd worker's accuracy on new tasks based on his accuracy on past  tasks. Such prediction provides a foundation for identifying the best workers to route work to in order to maximize accuracy on the new task. Our key insight is to model similarity of past tasks to the target task such that past task accuracies can be optimally integrated to predict target task accuracy. We describe two matrix factorization (MF) approaches from collaborative filtering which not only exploit such task similarity, but are known to be robust to sparse data. Experiments on synthetic and real-world datasets provide feasibility assessment and comparative evaluation of MF approaches vs.\ two baseline methods. Across a range of data scales and task similarity conditions, we evaluate: 1) prediction error over all workers; and 2) how well each method predicts the best workers to use for each task. Results show the benefit of task routing over random assignment, the strength of probabilistic MF over baseline methods, and the robustness of methods under different conditions. 
\end{abstract}

\section{Introduction}
\label{sec:intro}

\begin{figure*} [t]
\centering
\ifpdf
\includegraphics[height=32mm]{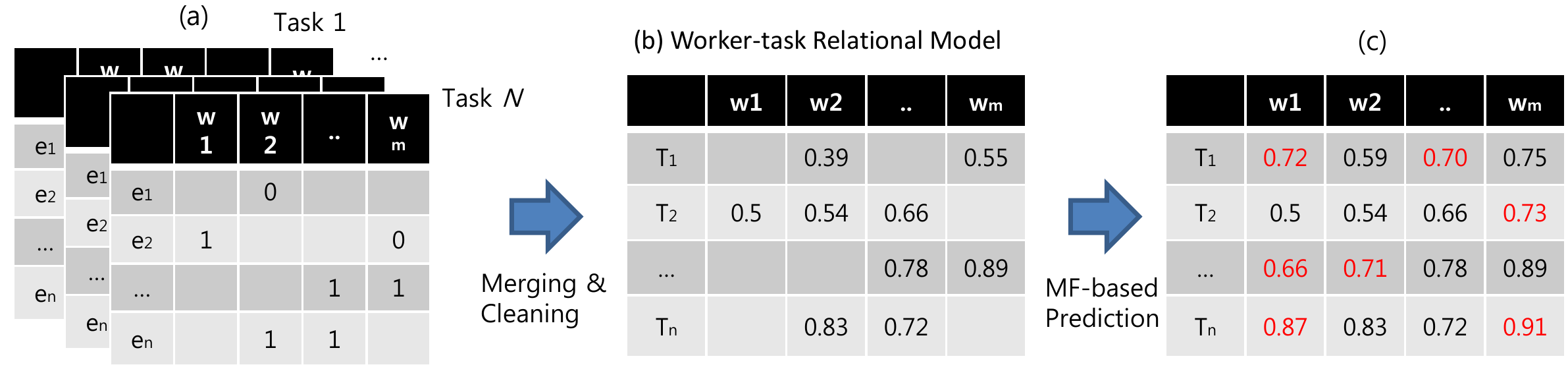}
\fi
\vspace{-10pt}
\caption{Matrix factorization (MF) prediction of crowd workers' accuracies using a worker-task matrix. (Left) A worker-example matrix contains labels from workers for each task. (Center) From the worker-example matrices, we measure each worker's accuracy vs.\ a ground truth sample, then merge all workers' accuracies into a single worker-task matrix. (Right) Unobserved workers' accuracies are predicted by MF (shown in red), and these predictions allow us to tailor individual work assignments to workers predicted to perform well for a given task.}
\label{fig:mf_process}
\vspace{-10pt}
\end{figure*}

Crowdsourcing is quickly changing data collection practices in both industry and data-driven research areas such as natural language processing~\cite{Snow2008}, computer vision~\cite{Vijayanarasimhan2011}, and information retrieval~\cite{Law2011a}. However, quality concerns persist, especially with rudimentary crowdsourcing platforms like Amazon's Mechanical Turk (MTurk) where factors such as anonymity, piece-work pay, and limited worker interaction can contribute to poor quality crowd work.  Various statistical approaches have been proposed for mitigating these issues, such as by aggregating inputs from multiple workers~\cite{Snow2008} or filtering out inaccurate workers~\cite{ramesh2012identifying}. 

MTurk's default use case assumes workers self-select tasks.  In principle, this is often a good idea. For example, using an in-house labor platform, Law et al.\ found higher quality work when workers were allowed to self-select work rather than be assigned it arbitrarily.  However, an important assumption here is that workers are able to effectively search and find tasks of interest. In fact, prior work has shown that limited task search capabilities on MTurk in practice often lead to task selection being more random than one might otherwise expect~\cite{Chilton2010}. Moreover, lack of support for task routing in MTurk's default setup has led to a dearth of research in this area. Nonetheless, 
like spam filtering, the promise of work filtering / tailored work assignments is to better match workers to work for which they are best suited, with potential to increase work quality and satisfaction and reduce inefficiency of task selection. 

We  describe methods to predict a crowd worker's accuracy on new tasks based on his accuracy on past tasks. Such prediction provides a foundation for identifying the best workers to route work to in order to maximize accuracy on the new task. Note our methods do {\em not} require example feature representations and so are broadly applicable across crowdsourcing tasks. Our key insight, based on preliminary analysis on our MTurk data (Section~\ref{sec:real_data}), suggests cross-task worker accuracies being correlated based on task similarity. Intuitively, more similar tasks should yield more similar worker accuracy across tasks. Of course, ``spammers'' may still perform uncorrelated, inaccurate work across tasks. By modeling similarity to past tasks, work history can be better integrated to predict new task accuracy. 

Critically, note that such prediction cannot be done for one-shot or small scale data collection, where there is no work history, or where workers completely little work before leaving and never returning. While many academic studies have tended to report low rates of worker retention across tasks and within-task completion, we posit this may reflect community sampling bias of one-off, infrequent academic studies. In contrast, commercial crowd work offers large volume and repetition that allows workers to amortize time to spent learning a task, returning to tasks for which they are already familiar for greater retention and completion rates. 

We describe two approaches to predict worker accuracies based on matrix factorization (MF), widely used in collaborative filtering problems to predict missing values in a matrix using low-rank feature vectors~\cite{Koren2009}. To predict unobserved workers' performance on a new task, we construct a worker-task matrix, where entries reflect a worker's observed accuracy on past tasks, evaluated against some sample of ground truth data (Figure~\ref{fig:mf_process}). We investigate two well-known MF models: Singular Value Decomposition (SVD) and Probabilistic Matrix Factorization (PMF)~\cite{Ruslan2008}. Prior work~\cite{Lee2012_CF} describes the two MF's tradeoffs for general recommendation systems. We revisit such questions by investigating these issues in our setting of crowd worker accuracy prediction.


Experiments on synthetic and real-world datasets provide feasibility assessment and comparative evaluation of MF approaches vs.\ two baseline methods. Across a range of data scales and task similarity conditions, we evaluate methods in two ways: RMSE prediction error over all workers, and average accuracy of the top \textit{k} ranked workers according to each prediction method.

\begin{description}
\item[RQ1: MF-based Prediction Accuracy]{\em How does MF prediction performance vary as a function of task similarity, matrix size, and matrix density for predicting worker accuracies across tasks? How feasible and robust is it to challenging conditions?}
\item[RQ2: Finding Top Workers]{\em Does task routing to predicted top $k$ workers outperform random assignment? If so, by what degree, and how do proposed MF methods (PMF and SVD) compare vs.\ simpler baseline methods (average and weighted average)?}
\item[RQ3: The Effect of Spammers]{How robust is MF-based task routing to the existence of spammers?}
\end{description}
Results show the benefit of task routing over random assignment, the strength of probabilistic MF over baselines, and robustness of methods to different conditions. 



\section{Related Work}



MTurk's standard method of task self-selection has led to relatively few studies on task routing to better match workers to tasks, though work considered task assignment in other venues, such as Wikipedia~\cite{cosley2007suggestbot}. 
Others have studied the cooperative refinement and task routing among on-line agents with regard to prediction tasks~\cite{KamarEce;HackerSeverin;Horovitz2012}. Bernstein at al.~\cite{Bernstein2012} investigate task routing in terms of real-time crowdsourcing. Though informative, these studies do not address finding strong candidates for a particular task from a task requester's viewpoint. Other work on task markets chains together of different worker competencies for problem solving~\cite{shahaf2010generalized}.

Karger et al.~\cite{Karger2011a} present a task assignment model based on random graph generation and a message-passing inference algorithm, in order to route tasks to crowd workers under homogeneous labeling tasks. Ho et al.~\cite{Ho2013} attempt to generalize this model to allow heterogeneous tasks by applying on-line primal-dual techniques. However, neither study answers the question of how to predict the unobserved workers' performance, which is critical to task routing in practice. Zhang et al.~\cite{zhang2012task} consider a related task routing task that seeks to engage people or automated agents to both contribute solutions and route tasks onward. 

Jung and Lease~\cite{Jung2012-hcomp} study MF methods to improve the quality of crowdsourced labels, using a PMF model to infer unobserved labels in order to reduce the bias of the existing labels. They do not consider task routing. Yi et al.\ investigate matrix completion for crowdclustering, and more recently inferring user preferences~\cite{yi2012crowdclustering,Yi13}. Yuen et al.~\cite{Yuen2012} consider workers' task selection preferences and propose a task recommendation model based on PMF. However, they motivated their approach on conceptual grounds and did not evaluate it. Most recently, Kolobov et al.~\cite{Kolobov13} investigate task routing of multiple tasks across a common pool of workers.

\section{Worker-Task Matrix Factorization}
\label{sec:mf_model}

Matrix factorization (MF) has been studied for effectively recommending an item for a user in an online marketplace and advertisement. Our intuition is that finding strong candidates for a specific task is very similar to the recommendation of items for a specific user in collaborative filtering~\cite{Jung2012-hcomp}. In addition, latent features should capture how a worker successfully makes a label for a specific task. For example, two workers would achieve high accuracy for a task type if they both have similar amounts of domain knowledge for this task type. If we can discover these latent features, we should be able to predict workers' accuracies by task. Given a partially observed worker-task matrix, we aim to predict unobserved workers' accuracies (e.g., on new tasks) so that we might route work optimally, or recruit or exclude particular workers for a given new task.


{\bf Singular Value Decomposition (SVD)} 
seeks an approximation matrix $\hat{R} = W^{T} T$ of the given rank which minimizes the sum-squared distance between the original matrix $R$ and $\hat{R}$. It has a critical drawback: it is 
undefined for incomplete matrices. Thus, it does not solve the problem of sparseness in the given $R$.

{\bf Probabilistic Matrix Factorization (PMF)} 
was introduced by Ruslan~\cite{Ruslan2008} and has demonstrated excellent performance in the Netflix challenge. We have {\it M} crowd workers, {\it N} tasks, and a worker-task matrix $R$ in which $R_{ij}$ indicates the accuracy of worker $i$ for task $j$. Let $W \in \mathbb{R}^{D*M}$ and $T\in \mathbb{R}^{D*N}$ be latent feature matrices for workers and tasks, with column vectors $W_i$ and $T_j$ representing $D$-dimensional crowd worker-specific and task-specific latent feature vectors, respectively. The conditional probability distribution over the observed cells $R \in \mathbb{R}^{M *N}$ is given by Equation 2. Indicator $I_{ij}$ equals 1 {\em iff} worker {\it i}'s accuracy is measured over task {\it j}. We place zero-mean spherical Gaussian priors on worker and task feature vectors.
To estimate model parameters, we maximize the log-posterior over example and worker features with fixed hyper-parameters. Maximizing the posterior with respect to {\it W} and {\it T} is equivalent to minimizing squared error with L2 regularization:
\vspace{-10pt}
\begin{multline}
E = \frac{1}{2}\sum_{i=1}^{M} \sum_{j=1}^{N} I_{ij}(R_{ij} - W_{i}^{T}T_{j})^{2} \\
+\frac{\lambda_{W}}{2}\sum_{i=1}^{M}\Vert W_{i} \Vert^{2}_{Fro} +\frac{\lambda_{T}}{2}\sum_{j=1}^{N}\Vert T_{j} \Vert^{2}_{Fro}
\end{multline}
where $\lambda_{W} = \sigma_{W} / \sigma, \lambda_{T} = \sigma_{T} / \sigma$, and $\Vert\centerdot \Vert_{Fro}^{2}$ denotes the Frobenius Norm. We use gradient descent to find a local minimum of the objective for $W$ and $T$. Finally, we infer unobserved workers' accuracies in the worker-task matrix $R$ by the scalar product of {\it W} and {\it T}.

\section{Experiments}
\label{sec:Experiments}

\begin{figure} [t]
\centering
\ifpdf
\includegraphics[height=31mm]{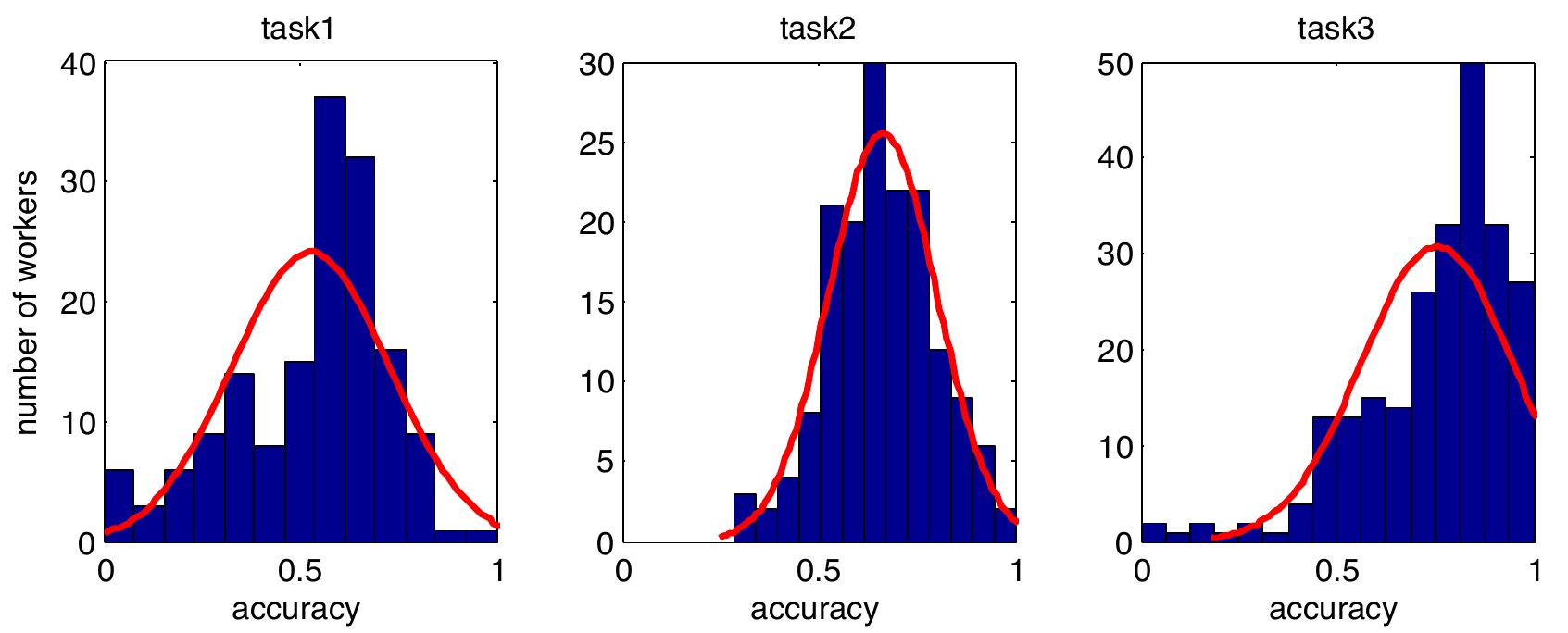}
\fi
\vspace{-10pt}
\caption{Histograms of Workers' Accuracy in three different tasks. All these three histograms show that workers' accuracy distribution follow a normal distribution.}
\label{fig:real_turk_histogram}
\vspace{-10pt}
\end{figure} 

\begin{figure} [t]
\centering
\ifpdf
\includegraphics[height=40mm]{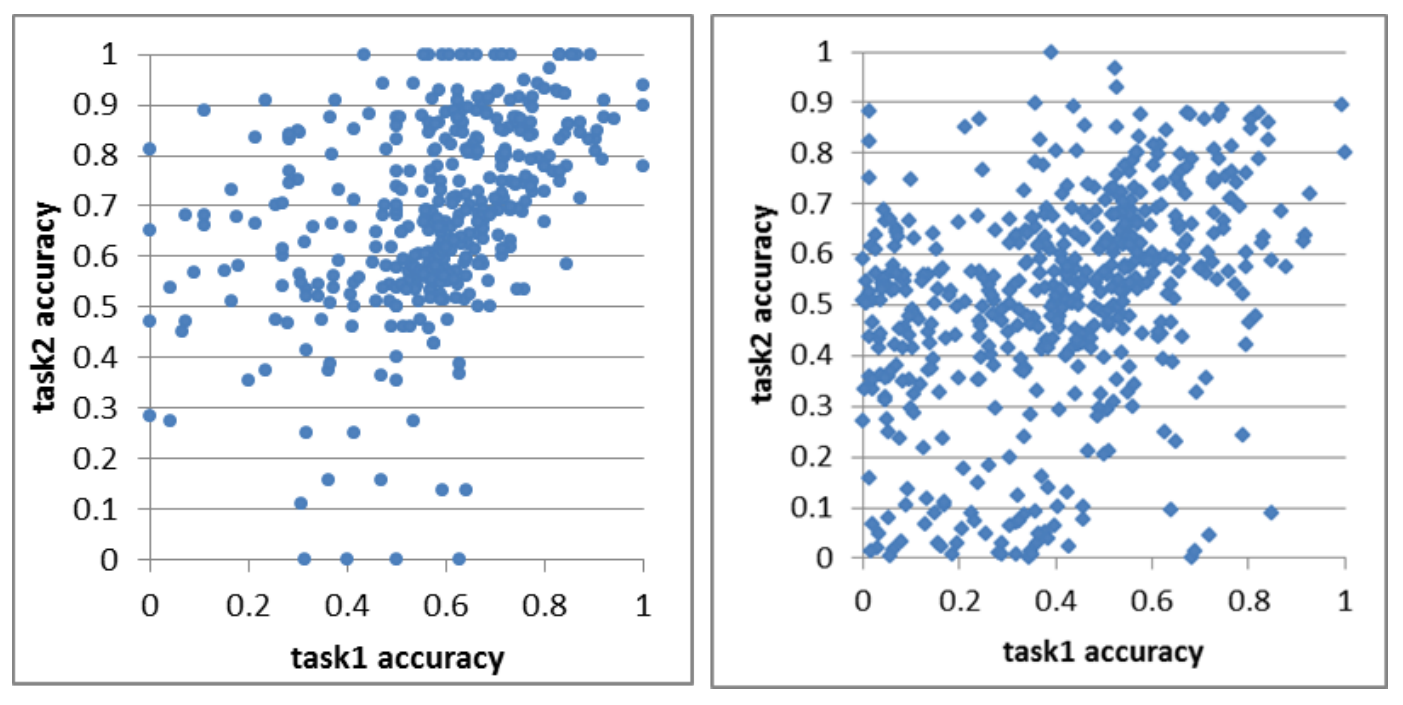}
\fi
\vspace{-10pt}
\caption{Scatter plots of accuracies across two tasks in (a) MTurk data and (b) synthetic data.}
\label{fig:scatter_tasks}
\vspace{-10pt}
\end{figure} 

 


Our first set of experiments use synthetic data, letting us carefully control a variety of experimental variables for detailed analysis.  Following this, we describe additional experiments with real crowd data to assess performance of methods for a specific case of actual operating conditions. We begin by describing these datasets.

Our generative model for synthetic data is based up the the MTurk dataset consisting of three tasks (Section~\ref{sec:real_data}). Figure~\ref{fig:real_turk_histogram} plots histograms showing the number of workers achieving various levels of accuracy in each of three MTurk tasks. These histograms show a strong normal tendency, which we quantify later via a Shapiro-Wilk test~\cite{Shapiro1965} (Table~\ref{table:summary_real_data}). In addition, Figure~\ref{fig:scatter_tasks} (left) plots average worker accuracy for task 1 vs.\ task2 and shows strong correlation at high accuracies (similar plots for task 1 vs.\ task 3 and task 2 vs.\ task 3 are not shown). In other words, the best workers appear to be fairly accurate across tasks, whereas other workers show less correlation across tasks, perhaps tending toward uncorrelated accuracies as average accuracy across tasks decreases (e.g., the increasing prevalence of ``spammers'' as we consider lower average accuracies). These observations suggest correlated worker accuracies across tasks, which might be reasonably well-fit by a multivariate normal distribution with appropriate covariance. We further develop this idea below.

\subsection{Synthetic Data.}
\label{sec:data_gen}

\begin{algorithm}
\caption{Generative model for synthetic data}
\label{alg:data_gen}
\begin{algorithmic}[1]
\Require $taskSimilarity,numWorkers,numTasks$
\State $\sigma = matrix(taskSimilarity,numTasks,numTasks)$
\State $m = multivarNormal(numWorkers,\mu = 0.5,\sigma)$
\ForAll {$t \in [1:numTasks]$}
	\State $accuracy[1:numWorkers] = m[t]$
	\ForAll {$j \in [0:0.8]$ by $0.1$}
		\State $strata = [j, j+0.1]$
		\State  $workers = findWorkers(accuracy,strata)$
		\State $\alpha = (0.8-j)*10$ \Comment $\alpha \in \{80,70,\ldots,0\}$
		\State $subset = getRandomSample(workers,\alpha\%)$
		\ForAll {$worker \in subset$} 
			\State $max = $0.9/(9-j*10) \Comment $max \in \frac{0.9}{[9,8,7,\ldots,1]}$
			\State $accuracy[worker] = uniform[0,max]$
		\EndFor
	\EndFor
\EndFor
\Ensure $m$
\end{algorithmic}
\end{algorithm}

Our model of worker behavior makes three key assumptions. Firstly, we model workers' accuracies as following a normal distribution N($\mu$, $\sigma$), based on our real crowd accuracy histograms (Figure~\ref{fig:real_turk_histogram}) and supported by a Shapiro-Wilk test~\cite{Shapiro1965} confirming high normality (Table~\ref{table:summary_real_data}). Secondly, we assume worker accuracy is correlated across tasks in proportion to task similarity. In addition to knowing people naturally exhibit varying expertise/skill across tasks, we also observe this in our real crowd data (correlation in Figure~\ref{fig:scatter_tasks}'s left plot). Thirdly, we posit the existence of ``spammers'' who exhibit low accuracy, uncorrelated across tasks, due to any number of factors, such as fatigue, low language competency, negligence, etc. Because better workers are less likely to be spammers (by definition), we expect fewer spammers when average worker accuracy is high, and the ratio increasingly shifting toward more spammers at lower accuracies.

Our generative model for synthetic data (Algorithm~\ref{alg:data_gen}) takes three parameters as input which we vary as experimental variables: task similarity $s$, number of tasks $t$, and number of workers $w$. We sample a multivariate normal distribution of accuracies (per-task average accuracy), providing correlation across tasks by specifying covariance matrix $\sigma$ filled with $s$ ({\tt rmvnorm} function from {\tt mvtnorm} library in $R$). This ``optimistic'' distribution is free of spammers and reflects an idealized model of correlated accuracies. Next, we introduce spammers to produce an alternative ``pessimistic'' distribution by transforming a percentage of the idealized workers into spammers. To accomplish this, we first group workers by distributional strata over average accuracy across tasks (using a sliding window of size $0.1$). We then randomly sample a percentage of diligent workers from each strata and transform them into spammers by replacing their idealized per-task accuracy on each task with an accuracy sampled uniformly at random. 

Note that as a function of strata, we must decide (a) what percentage of workers to transform, and (b) the maximum accuracy of the interval from which to sample spammer accuracies. While our choice of functions here for (a) and (b) are relatively ad hoc, they enforce our principle above of finding fewer spammers as worker accuracy increases. We argue strengths of this model include: (i) its full description for reproducibility by others~\cite{paritosh2012human}; (ii) its implementation of over-arching modeling assumptions in some reasonable way; and (iii) an explicit discussion of how more accurate simulation might be achieved by further analysis and characterization of real crowd data properties. 


\begin{table}
\centering
\begin{tabular}[h]{|c| c| c|c |}
\hline
{\bf \!\!Tasks} & {\bf Workers} &{\bf Accuracy} & {\bf Normality Test\!\!}\\ 
\hline
{\bf \!\!Task1} &206 & 0.676 & 0.9471 *\\ 
{\bf \!\!Task2} &384 & 0.599 & 0.99 *\\ 
{\bf \!\!Task3} &167 & 0.491 & 0.90 *\\ 
\hline
{\bf \!\!All} &443 & 0.596 & \\ 
\hline
\end{tabular}
\caption{Attributes of MTurk data with three tasks. Task similarity $s$ ranges from [0.545:0.719]. For all tasks, distribution of per-task worker accuracies follow a normal distribution with high statistical significance ($p<0.01$) under the Shapiro-Wilk test~\cite{Shapiro1965}.}
\label{table:summary_real_data}
\vspace{-20pt}
\end{table}

\subsection{MTurk Data.}
\label{sec:real_data}

We experiment with a commercial MTurk dataset which includes three different text processing tasks. Only workers with high approval ratings were allowed to perform work. While we cannot further describe the tasks or their similarity, we will show task similarity can be inferred automatically and varies. Moreover, asking a person to articulate a similarity score for two tasks is not a natural activity and expected to be difficult anyway, without experience or guidelines. As such, the ability to automatically infer task similarities without any reliance on manual tweaking is a strength of our approach.  

We also note ours is a retrospective study: workers self-selected what work to perform during actual data collection, and we simulate task assignment now after the fact. While such retrospective assignment methodology represents a common study design in active learning and community question answering research, we nonetheless method the usual limitations of such a study design.  In practice, we would have to contend with workers not performing tasks assigned to them, or that accuracy could diminish when people are assigned work instead of self-selecting that which most interests them.

Because a central goal of our work is to identify strong workers to whom work could be usefully assigned, there is little value in modeling workers who only briefly try out a task without completing much work. Consequently, we exclude workers who completed fewer than ten examples per task (37.4\% of workers retained for Task1, 43.2\% for Task2, and 30.8\% for Task 3). While modeling these other workers could be interesting in multiple respects (e.g., to study these ``flighty'' workers, or estimating worker accuracies from extremely sparse observations), we leave this for future work. 

Table~\ref{table:summary_real_data} characterizes remaining data for each task. 443 workers completed at least 10 examples on all 3 tasks, and 43\% of the (443 x 3) worker-task matrix observed. While prior crowdsourcing studies have tended to report lower rates of worker retention across tasks and within-task completion, the difference seen in our data may be indicative of a sampling bias between what is observed in one-off, infrequent academic studies vs.\ what is seen in commercial crowd work, where volume and repetition of familiar tasks lead to greater worker retention and completion rates. Greater study of commercial crowd data to assess such potential bias will be another interesting direction for future work.


\subsubsection{Metrics.}
\label{sec:metrics}

We report root-mean squared error (RMSE) and the mean accuracy of the top \textit{k} workers (\ma) according to some ranking function:
\vspace{-5pt}
\begin{equation*}
RMSE =\sqrt{\frac{\sum_{i,j} {(\hat{a}_{i,j} - a_{i,j})^{2}}}{n}}~~,~~ 
MA_k = \frac{1}{k}{\sum_{i=1}^{k}}{a_{i,j}}
\end{equation*}
%
where $\hat{a}_{i,j}$ and ${a}_{i,j}$ respectively denote the predicted vs.\ observed worker accuracy for worker $i$ on task $j$. 


While task similarity $s$ could be operationalized in many ways, in this study we adopt Pearson's correlation coefficient $r$ (i.e., $s=r$ in our study, but we preserve distinct notation of $s$ for generality). Following standard practice~\cite{Cohen1988}, we distinguish four specific levels of correlation in our experiments: weak ($r$=0.1), medium ($r$=0.4), strong ($r$=0.7), and very strong ($r$=0.9).

\subsubsection{PMF Model Parameters.}

MF infers unobserved, low dimensional factors (feature vectors) to explain observed workers' accuracies. We want to find an optimal dimensionality $D$ for these latent feature vectors. While larger $D$ is expected to yield more accurate prediction, it comes at a computational cost. Thus, it is common to optimize $D$ on development data. This development data is also used to optimize PMF regularization parameters $\lambda_{w}$ and $\lambda_{t}$ and learning rate $\epsilon$. Unlike our PMF formulation, SVD is parameter-free.

$\lambda_{w}$ and $\lambda_{t}$ are selected via 5-fold cross-validation, using {\bf only} 20\% of data for training and leaving 80\% for testing. 
For simplicity, we try only three settings each for $\lambda_{w}$ and $\lambda_{t} \in \{10^{-1}, 10^{-2}, 10^{-3}\}$, ultimately selecting $\lambda_{w} = \lambda_{t} = 0.01$. We use $\epsilon=0.005$ without tuning.


\subsubsection{Baselines.}

One goal of our work is to assess the potential benefit of task routing at all (i.e., any approach) vs.\ just assigning tasks arbitrarily (i.e., random task assignment). While random assignment represents a very simplified model of how MTurk workers self-select tasks today, prior work has shown that MTurk's limited task search capabilities lead to task selection often being more random than one might expect otherwise~\cite{Chilton2010}. Thus for task routing experiments, random assignment provides our simplest, uninformed baseline.

A related but distinct goal of our work is to assess relative improvement of specific approaches proposed in Section~\ref{sec:mf_model} (SVD and PMF) vs.\ alternative, simpler but reasonable baseline methods one might utilize for automatic task assignment. 

Our first such informed baseline is to simply predict a worker's accuracy on a new task by his average accuracy on other tasks. A limitation of this baseline, however, is that it completely ignores task similarity. Moreover, since our generative model for synthetic data explicitly generates worker accuracies correlated across tasks by similarity (Section~\ref{sec:data_gen}), we expect a baseline exploiting this additional information should perform better. Our second informed baseline therefore computes an expectation (weighted average) across tasks based on task similarity rather than a simple average. 


\begin{figure*} [t]
\centering
\ifpdf
\includegraphics[width=170mm]{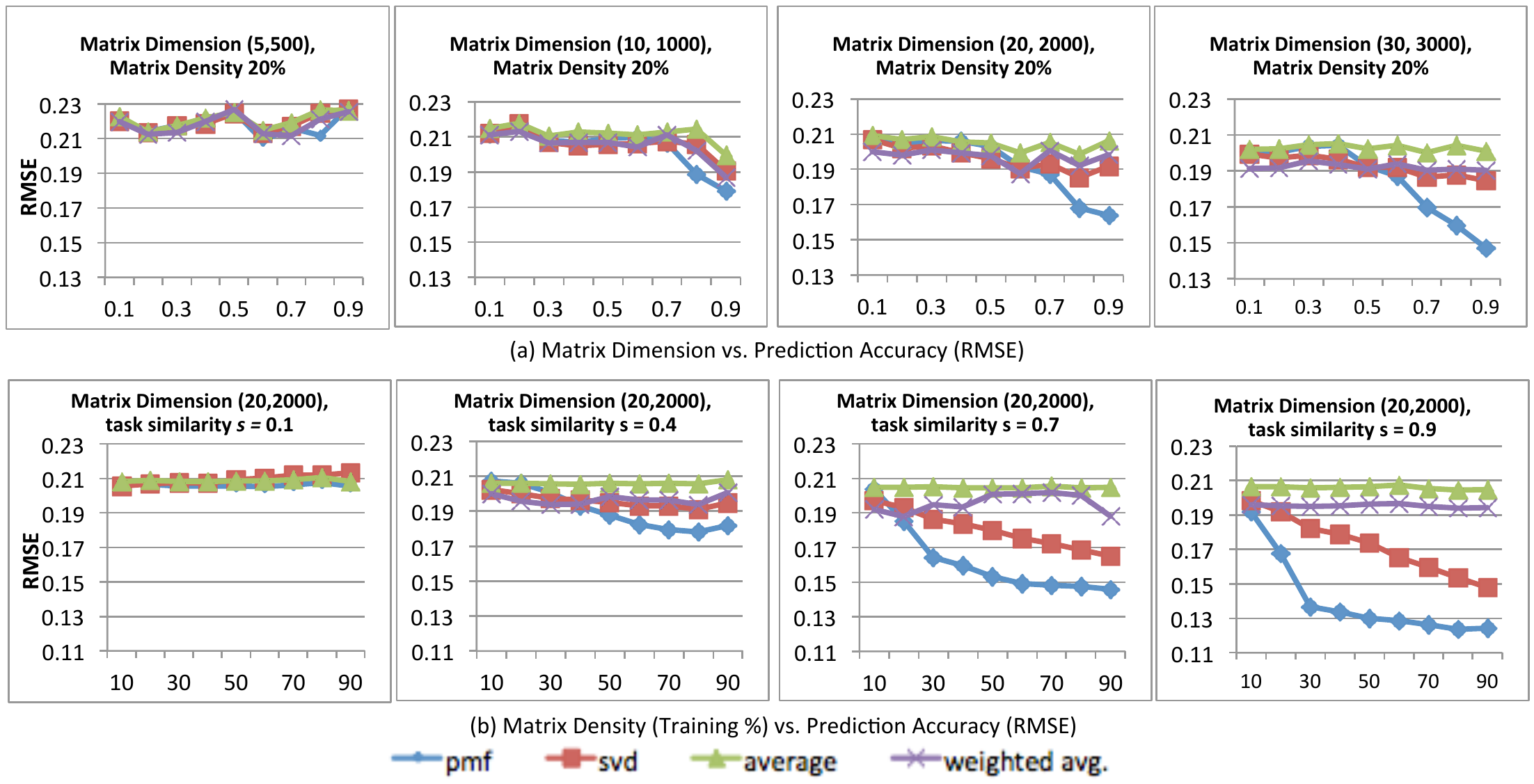}
\fi
\vspace{-10pt}
\caption{(a) Matrix dimension vs.\ prediction accuracy and (b) Matrix Density (Training \%) vs.\ prediction accuracy. In (a), the x-axis shows the increase of task similarity ranging from 0.1 to 0.9 by 0.1. In order to investigate the effect of matrix dimension, we evaluate the prediction accuracy by increasing the number of tasks ($t$) and the dimensionality of feature vectors ($d = t-1$). Matrix density is fixed as 20\%. In (b), the x-axis shows the increase of matrix density from 10\% to 90\% by 10\%. In addition, we increase task similarity from 0.1 to 0.9 while fixing matrix dimension as 20 by 2,000 and the dimensionality of feature vectors ($d = 19$). 
}
\label{fig:task_similarity}
\vspace{-10pt}
\end{figure*}

\subsection{Synthetic Dataset Experiments}
\label{sec:eval_results}

\subsubsection{Experiment 1 (RQ1): MF-based Prediction Accuracy.}

%

{\em To what extent do task similarity, matrix size, and matrix density influence MF-based prediction of crowd worker accuracies for varying task similarity?} We first measure RMSE of PMF and SVD methods vs.\ baseline methods for task similarity $s$ (Pearson's correlation $r$) $\in \{0.1,0.2,\ldots,0.9\}$, averaging 1,000 simulation trials for each setting of $s$. Results in Figure~\ref{fig:task_similarity} (a) and (b) show the RMSE achieved along the y-axis.

Our first set of experiments (Figure~\ref{fig:task_similarity}(a)) evaluate our ability to effectively predict worker accuracies across varying task similarity $s$, number of tasks, and number of workers. We vary the number of tasks from 5 to 30 (by 5) and the number of workers from 100 to 3000 (by 500). Four configurations of varying number of tasks $t$ vs.\ number of workers are shown: (5,500), (10,1000), (20, 2000), and (30,3000). Task similarity $s$ is varied along the x-axis. Matrix density is fixed at 20\% and dimensionality $D$ is set by $D=t-1$. 

Our second set of experiments (Figure~\ref{fig:task_similarity}(b)) evaluate our ability to effectively predict worker accuracies under varying task similarity $s$ and matrix density. Each of the four plots corresponds to a different task similarity $s \in \{0.1, 0.4, 0.7, 0.9\}$, matching the levels of Pearson correlation $r$ from {\em weak} to {\em very strong} as defined in Section~\ref{sec:metrics}. Matrix density varies along the x-axis (from 10-90\%). We fix the number of tasks ($t=20$), dimensionality ($D=t-1$), and the number of workers (2000). 

\begin{figure*} [t]
\centering
\ifpdf
\includegraphics[width=170mm]{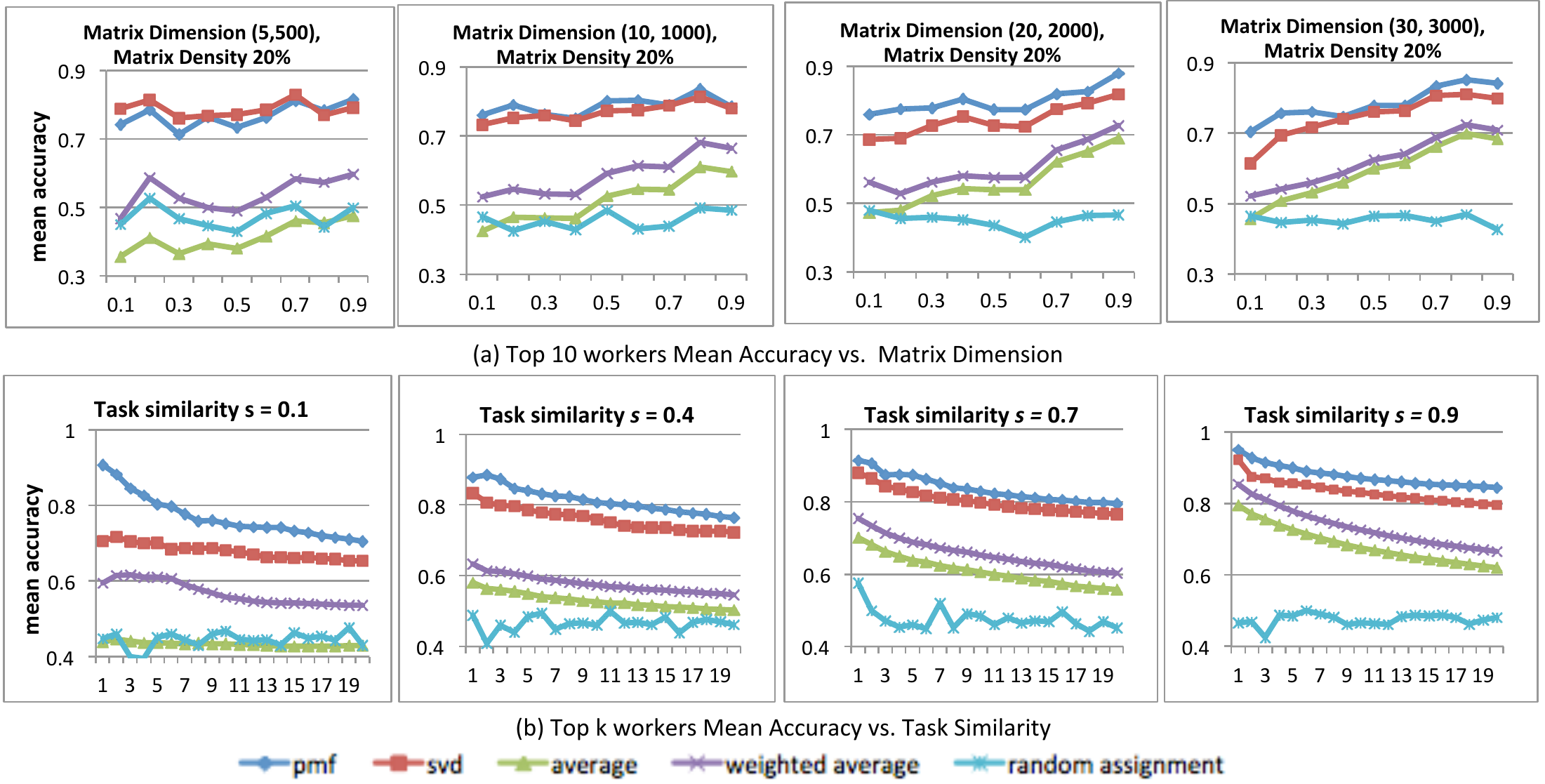}
\fi
\vspace{-10pt}
\caption{(a) Matrix dimension vs.\ top \textit{10} workers' MA$_{10}$. The x-axis shows the increase of task similarity [0.1:0.9] and matrix dimension is increased across plots. (b) Task similarity vs.\ top \textit{k} workers' mean accuracy {\ma}. The x-axis shows the increase of top \textit{k} workers from 1 to 20. We evaluate {\ma} across increasing task similarity (0.1, 0.4, 0.7, 0.9). 15 tasks and 1,500 workers are used. Both (a) and (b) fix matrix density at 20\%.}
\label{fig:ma_exp}
\vspace{-15pt}
\end{figure*}

{\bf Task Similarity.} 
{\em How similar do tasks need to be before we can make effective predictions?  How much do predictions improve with greater task similarity?} With only weak similarity $s=0.1$ (left-most point in all 4 plots in (a), left-most plot in (b)), RMSE of 0.21-0.23 can still be achieved, though this is far below the best RMSE$<$0.13 observed with very high task similarity.  Baselines seem sufficient, without need for MF methods.

As task similarity increases (across x-axis in (a) plots, and across plots in (b)), it tends to enable better prediction of worker accuracies across tasks, as expected. Moreover, we see PMF predictions improve by exploiting greater task similarity: both as the number of tasks and workers increase (a), or as matrix density increases (b). In contrast, SVD performs comparably to baselines in all four of the plots in (a); it does not benefit from increased task similarity even as the matrix size grows. While in (b) plots we do see SVD perform much better as $s$ increases with greater density, PMF still outperforms SVM by a wide margin.  

{\bf Matrix Size.} {\em How effectively can we predict worker accuracies when only small worker-task matrices (relatively few workers or tasks) are observed?  At what point (if any) does accuracy prediction break due as the matrix becomes too small?  How much does prediction improve with larger matrix size?}

Figure~\ref{fig:task_similarity}(a) shows that in the smallest case of 5 tasks and 500 workers, RMSE of 0.21-0.23 can still be achieved. Moreover, baseline methods seem sufficient in this case (comparable to MF methods). However, this is far below the sub-0.15 RMSE achieved by PMF with larger matrix sizes. As noted above, we see PMF capitalize on increasing task similarity $s$ with larger matrices while SVD does not.

{\bf The Effect of Matrix Density/Sparseness.} Prior crowdsourcing studies often report workers are often transient and rarely complete all tasks available. {\em What matrix density is needed to support effective worker accuracy predictions across tasks? How does prediction accuracy improve with increasing density?} 


Figure~\ref{fig:task_similarity}(b) shows across plots, RMSE lies in [0.19-0.21] when density is only 10\%. As before, baseline methods seem sufficient when density is so low, with no improvement from MF methods. However, as noted above, both SVD and PMF improves dramatically vs. baselines with greater density, with PMF consistently dominating SVD across (b) plots. The best case RMSE$<$0.13 is observed under ideal settings of an extremely dense matrix and large worker-task matrix.


{\bf RQ1 Summary.} Under worst case conditions, results show worker accuracies can be predicted with RMSE$\in[0.19-0.23]$, and that even simple averaging suffices is comparable to MF methods. However, as tasks exhibit even medium task similarity ($s=0.4$) significant gains are possible, and especially with steep with high correlation ($s=0.7$). Similar wins occur as matrix size increases, and with steep RMSE decreases beginning with matrix densities around only 30\%.  PMF consistently outperforms SVD in all cases where sufficient data exists to outperform baselines.

\subsubsection{Experiment 2 (RQ2): Finding Top Workers.}

RQ1 investigated our ability to predict worker accuracies across tasks in terms of RMSE.  For task routing, however, what we would really like to do is identify the top $k$ workers for a given task and route work to them; we do not care about specific accuracies, nor do we care about workers below this top $k$. We would like to know if routing work to some top $k$ workers (via any method) outperforms random assignment by some margin. If so, we would like to evaluate the relative benefit of proposed MF methods (PMF and SVD) vs.\ the simpler baseline methods (average and weighted average).

To answer these questions, we use some method (informed or random) to identify a set of $k$ workers, and we measure the average accuracy of these $k$ workers to compute {\ma} for each method. As in earlier experiments, informed methods use limited training set knowledge (20\%) of a worker's accuracy on $t-1$ other tasks to predict the worker's accuracy on held out task $t$, and we perform this prediction in round-robin fashion across tasks, with each task held out in turn and accuracy of workers on it predicted from the other tasks. In experiments here, after computing the average worker across $k$ workers for each task, we then average this average accuracy across tasks to arrive at {\ma} for each method.  We vary $k$ in these experiments.


As in Figure~\ref{fig:task_similarity}, Figure~\ref{fig:ma_exp} shows two sets of experiments (a) and (b) of four plots each. The first figure (a) shows matrix dimension vs.\ top \textit{10} workers' MA$_{10}$. The x-axis shows increasing task similarity [0.1:0.9]; matrix dimension is increased across plots. The figure (b) shows task similarity vs.\ top \textit{k} workers' mean accuracy {\ma}. The x-axis shows increasing $k\in[1:20]$ of top \textit{k} workers. We evaluate {\ma} across increasing task similarity (0.1, 0.4, 0.7, 0.9). 15 tasks and 1,500 workers are used. Both (a) and (b) fix matrix density at 20\%.

Results here are simpler and easier to understand than those in RQ1: informed assignment methods consistently outperform random assignment, MF methods outperform baselines, and PMF outperforms SVD. Overall, increase in task similarity brings increase in MA$_{10}$ (Figure~\ref{fig:task_similarity} (a)). Increase in matrix dimension also leads to a better performance in terms of finding strong candidates. When matrix dimension is very small (5,500), we see some fluctuations due to spammers.  


\subsubsection{Prediction With or Without Spammers.}
\label{sec:eval_random_workers}

\begin{figure} [t]
\centering
\ifpdf
\includegraphics[width=87mm]{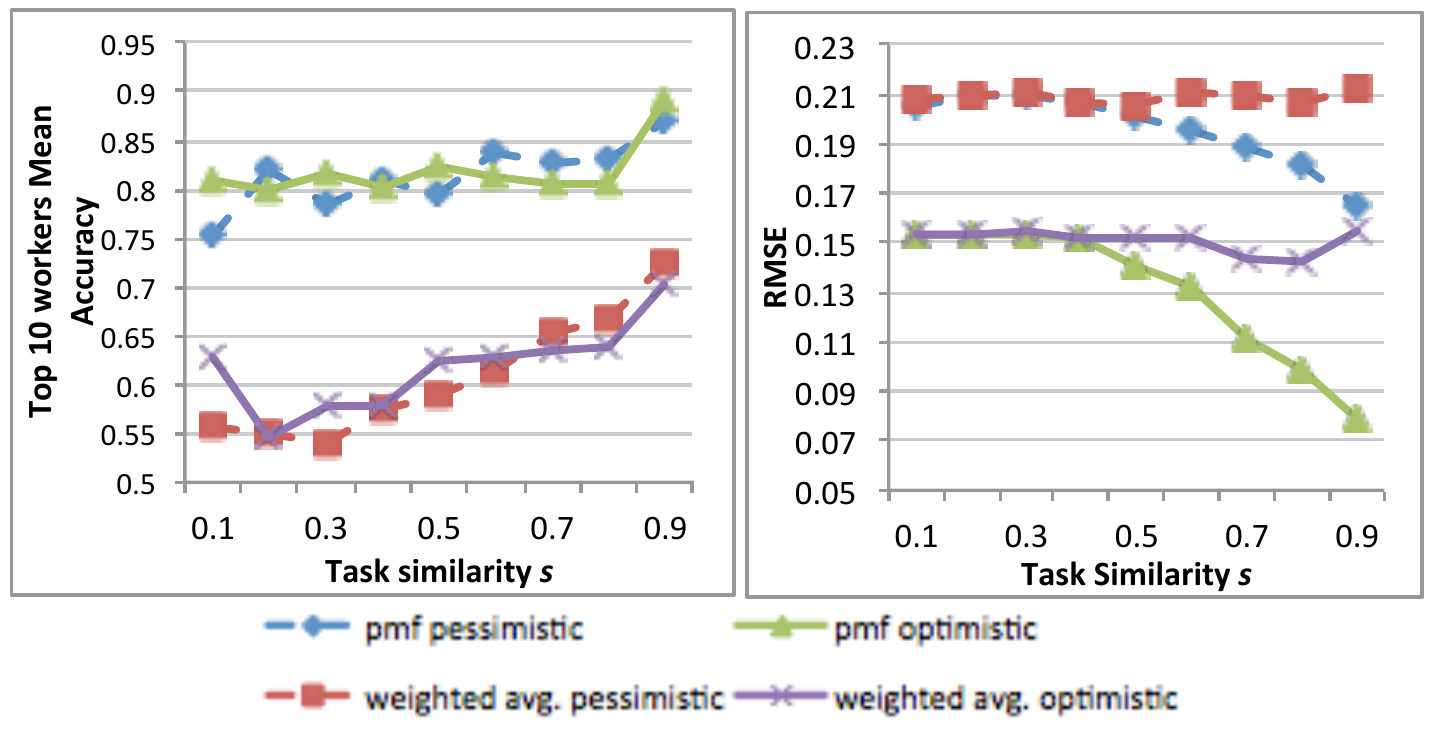}
\fi
\vspace{-20pt}
\caption{Prediction of worker accuracies when spammers are absent or present (optimistic or pessimistic controlled synthetic data experiment), measuring the top-$k$ workers' mean accuracy {\ma} (left plot, bigger is better) and RMSE over all workers (right plot, lower is better). Task similarity is varied along the x-axis. We compare PMF vs.\ the weighted average baseline.}
\label{fig:random_effect}
\vspace{-20pt}
\end{figure}

As discussed in Section~\ref{sec:data_gen}, spammers can be expected to degrade predictions of worker accuracies across tasks since such workers' accuracies across tasks are uncorrelated (by definition). {\em How robust are cross-task accuracy prediction methods to the presence of spammers?}


Figure~\ref{fig:random_effect} compares the PMF vs.\ the best informed baseline (weighted average) in the presence or absence of spammers. For the experiment, we use the fixed number of workers (1,500) and tasks(5). In regard to {\ma} (left plot), PMF vs.\ baseline results are basically equivalent with or without spammers, with PMF dominating weighted average as both trend slightly up with greater task similarity. As for RMSE (right plot), we also see PMF dominate the baseline across task similarity settings, with or without spammers. However, for RMSE we see that with or without spammers, baseline performance is flat (no error reduction from increasing task similarity), whereas PMF prediction error decreases with greater task similarity. Trend lines for each method are remarkably parallel with or without spammers, with an offset of about 0.05 RMSE absolute. 

We understand these findings as follows. To predict the top-$k$ workers for a given task, we focus on only the most accurate workers, which are easy to distinguish from spammers on the basis of higher accuracies on observed tasks (training data). Top-$k$ prediction therefore seems intuitively robust to spammers.  RMSE, on the other hand, is trying to predict accuracies of all workers, thus is inherently sensitive to spammers.  Nevertheless, we see PMF is still able to achieve increasing RMSE reductions with greater task similarity, whereas average accuracy is not, as well as PMF dominating average accuracy for {\ma} prediction as well. 

\subsection{MTurk Data Experiments.}
\label{sec:real_exp}

Whereas our experiments with synthetic data allowed us to carefully control a variety of experimental variables for detailed analysis, we also want to evaluate methods with real crowd data reflecting a specific case of actual operating conditions in which methods could be applied. With three tasks, we fix dimensionality $D=3$.  

We first consider the incomplete matrix of 443 workers who worked on all three tasks (Table~\ref{table:summary_real_data}). Due to sparsity, we do not always have training data on two other tasks to predict the third, so we cannot compute average and weighted average baselines. We see PMF, SVD, and random assignment achieve 0.170, 0.170, and .317 RMSE, respectively.  For MA$_{10}$, they achieve 0.773, 0.609, and 0.381, respectively. We see random assignment perform much worse, and PMF outperforming SVD for MA$_{10}$, but not for RMSE.

\begin{table}
\resizebox{\columnwidth}{!}{%
\centering
\begin{tabular}[bp]{|c |c| c| c || c| c| c| c| }
\hline
	    &\multicolumn{3}{c||}{RMSE}  & \multicolumn{4}{c|}{MA$_{10}$} \\
\hline
{\bf Tasks} &{\bf PMF} &{\bf WAvg} &{\bf Avg} & {\bf PMF} & {\bf WAvg} & {\bf Avg} & {\bf Rand} \\ 
\hline
{\bf task1} & 0.227 & 0.278 & 0.289 & 0.901 & 0.834 & 0.825 & 0.684\\ 
{\bf task2} &0.209  & 0.266 & 0.273 & 0.932 & 0.834 & 0.821 & 0.688\\ 
{\bf task3} &0.197 & 0.257 & 0.267 & 0.946 & 0.712 & 0.712 & 0.687\\ 
\hline
{\bf All} &0.211 &0.267 &0.276 & 0.926 & 0.793 & 0.786 & 0.686\\ 
\hline
\end{tabular}
} 
\caption{Performance on MTurk data. The RMSE and mean accuracy {\ma} of the top $k=10$ workers are shown for PMF vs.\ the three baselines: weighted average, average, and random assignment.}
\label{table:summary_real_turkdata}
\vspace{-10pt}
\end{table}

\begin{figure} [t]
\centering
\ifpdf
\includegraphics[width=65mm]{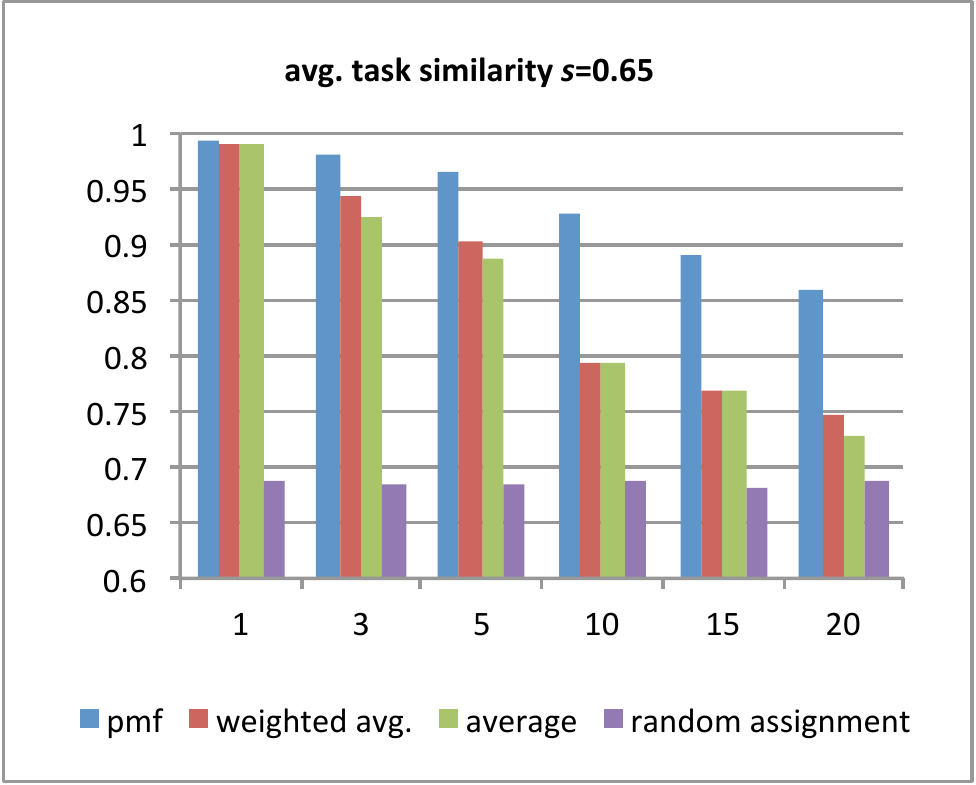}
\fi
\caption{Mean accuracy {\ma} of the top $k=10$ workers (averaged across tasks) as predicted by PMF, weighted average, and average, vs.\ random assignment.}
\label{fig:real_turk_ma}
\vspace{-20pt}
\end{figure}


Next we consider a smaller, denser matrix of the 54 workers who completed all examples for all tasks, allowing us to compute average and weighted average baselines. Figure~\ref{fig:real_turk_ma} shows {\ma} achieved (averaged across tasks) by different methods (informed PMF, weighted average, and average, vs.\ uninformed random assignment) for increasing $k \in \{1,3,5,10,15,20\}$. Average task similarity across each pair of tasks is estimated as $s=0.65$. Table~\ref{table:summary_real_turkdata} shows per task RMSE and {\ma} achieved by the same set of methods for fixed $k=10$.  

Overall, PMF is seen to dominate other methods except with $k=1$, where both informed baselines perform comparably. Interestingly, we do not see the weighted average offering benefit over the simple average here. We also observe that PMF and informed baselines all outperform uninformed random assignment by a wide margin, across settings. This supports our hypothesis of informed task assignment (by any means) being valuable over random assignment, and it is consistent with earlier findings on synthetic data. 

Task-averaged PMF RMSE in Table~\ref{table:summary_real_turkdata} of 0.21 appears consistent with synthetic experiments with 5 tasks, 500 workers, and $s=0.65$ (Figure~\ref{fig:task_similarity} (a)), where PMF achieved RMSE 0.23. Similarly, the trend of PMF {\ma} degradation with increasing $k$ in Figure~\ref{fig:real_turk_ma} (from roughly 0.98 to 0.86) appears consistent with Figure~\ref{fig:ma_exp} ($s=0.7$ plot) where {\ma} is seen to decrease comparably (from roughly 0.92 to 0.80).

\vspace{-5pt}
\section{Conclusion}

Task routing represents a relatively little explored and increasingly important area for future quality improvements in crowdsourcing. Many interesting questions remain open, such as modeling time-varying worker accuracies, due to fatigue or training effects, in determining appropriate task routing. Better ways to tackle sparse worker history data and longer term longitudinal worker studies could also be incredibly informative.


{\small {\bf Acknowledgments.} We thank the anonymous reviewers for their time and feedback. This work is supported in part by DARPA YFA Award N66001-12-1-4256, IMLS Early Career grant RE-04-13-0042-13, and NSF CAREER grant 1253413. Any opinions, findings, and conclusions or recommendations expressed by the authors do not express the views of any of the supporting funding agencies.}


\vspace{-5pt}
\bibliographystyle{siam}
{\small
\bibliography{jung_sdm14}
}
\end{document}